\documentclass[preprint,showpacs,preprintnumbers,amsmath,amssymb]{revtex4}

\usepackage{graphicx}
\usepackage{dcolumn}
\usepackage{bm}

\begin{document}


\title{Leptogenesis and Bi-Unitary Parametrization \\
of Neutrino Yukawa Matrix}

\author{Kim Siyeon}
\email{siyeon@phys.sinica.edu.tw}
\affiliation{%
Institute of Physics, Academia Sinica\\
Nankang, Taipei, Taiwan 11529
}%

\date{\today}

\begin{abstract}
\noindent We analyze the neutrino Yukawa matrix by considering
three constraints: the out-of-equilibrium condition of lepton
number violating process responsible for leptogenesis, the upper
bound of branching ratio of lepton flavor violating decay, and the
prediction of large mixing angles using the see-saw mechanism. In
a certain parametrization with bi-unitary transformation, it is
shown that the structure which satisfies the constraints can be
characterized by only seven types of Yukawa matrices. The
constraint of the branching ratio of LFV turns out as a redundant
one after applying other two constraints. We propose that this
parametrization can be the framework in which the CP asymmetry of
lepton number violating process can be predicted in terms of
observable neutrino parameters at low energy , if necessary, under
assumptions following from a theory with additional symmetries.
There is an appealing model of neutrino Yukawa matrix considering
the CP asymmetry for leptogenesis and the theoretical motivation
to reduce the number of free parameters.
\end{abstract}

\maketitle

\newcommand{\be}{\begin{equation}}
\newcommand{\ee}{\end{equation}}
\newcommand{\bea}{\begin{eqnarray}}
\newcommand{\eea}{\end{eqnarray}}
\newcommand{\yuk}{\mathcal{Y}}
\newcommand{\unir}{\mathbb{R}}
\newcommand{\unil}{\mathbb{L}}
\newcommand{\mns}{\mathbb{V}}
\newcommand{\br}{$BR$}

\section{Introduction: Leptogenesis}
\noindent%
The observed Baryon Asymmetry in Universe(BAU) can be explained
with a process that satisfies three conditions: Baryon number(B)
violation, Charge conjugation(C) violation and Charge conjugation
and Parity(CP) violation, and for the B-violating process to be
out-of-equilibrium \cite{Sakharov:dj}. Leptogenesis is a scenario
for BAU in which the initial lepton asymmetry is recycled into a
baryon asymmetry by the sphaleron process \cite{Kuzmin:1985mm}.
The initial lepton asymmetry can originate from loop-level
processes involving the Yukawa couplings of Majorana neutrinos
\cite{Fukugita:1986hr}. The baryon asymmetry $\Delta B$ can be
generated after sphaleron process in thermal equilibrium washes
$\Delta(B+L)$ out of $\Delta(B-L)+\Delta(B+L)$, where the
$\Delta(B-L)$ is equivalent to the initial lepton asymmetry
$-\Delta L$ \cite{Kuzmin:1985mm,Khlebnikov:sr,Harvey:1990qw}. The
constraints on chemical potentials of particles in thermal
equilibrium establish the relations among different asymmetries.
The amount of baryon asymmetry is described in terms of the ratio
of particle number density to entropy density, i.e., baryon number
density with respect to comoving volume element as
\begin{eqnarray}
    Y_B = {(n_B - n_{\overline{B}}) \over s},
\end{eqnarray}
with entropy $s$, the range of which is \cite{Hagiwara:fs}
\begin{eqnarray}
    Y_B \approx (0.6-1)\times 10^{-10}
    \label{aabaryon}
\end{eqnarray}
The baryon asymmetry is related to the lepton asymmetry
\cite{Harvey:1990qw,Kolb:vq} as
\bea %
Y_B = a Y_{B-L} = {a \over{a-1}} Y_L ,%
\eea %
where
\begin{eqnarray}
    a \equiv {{8 N_F + 4 N_H} \over {22 N_F + 13 N_H}},
\end{eqnarray}
for example, $a=28/79$ for the Standard Model(SM) with three
generations of fermions and a single Higgs doublet, $N_F = 3, N_H
= 1$, and $a=8/23$ if $N_F = 3, N_H = 2$ as in supersymmetric
models. Thus, the amount of the original lepton asymmetry should
be not less than $\sim 10^{-9}$ to provide necessary amount of
baryon asymmetry.

The generation of a lepton asymmetry also requires the
CP-asymmetry and out-of-equilibrium condition. The $Y_L$ is
explicitly parameterized by two factors, $\epsilon_i$, the size of
CP asymmetry, and $\kappa$, the dilution factor from washout
process.
\bea %
Y_L = {(n_L - n_{\overline{L}})\over s} = \kappa {\epsilon_i \over
g^*} \label{aalepto}
\eea %
where $g^*\simeq 110$ is the number of relativistic degree of
freedom. The $\epsilon_i$ is the magnitude of CP asymmetry in
decays of heavy Majorana neutrinos \cite{Kolb:qa,Luty:un},
\bea %
\epsilon_i
        &=&{{\Gamma (\nu_R \to \ell H)
            - \Gamma (\nu_R \to \ell^c H^c)}
        \over {\Gamma (\nu_R \to \ell H)
            + \Gamma (\nu_R \to \ell^c H^c)}},
\label{aacp}
\eea %
where $i$ is 1 to 3 for generation. When one generation of the
right neutrinos has a mass far below the masses for the other
generations, i.e., $M_1 \ll M_2, M_3$, the $\epsilon_i$ in
Eq.(\ref{aacp}) reduces to $\epsilon_1$ from the decay of $M_1$,
\bea %
    \epsilon_1 &=& {3\over {16\pi}}
            {1\over {({\yuk_N}^\dagger \yuk_N)_{11}}}
                \sum_{n\ne 1}{\rm Im}
            \left [({\yuk_N}^\dagger \yuk_N)_{1n}^2 \right ]
            {{M_1}\over {M_n}}\;,
\label{aacp1}
\eea %
where $\yuk_N$ is the matrix of Yukawa couplings in a weak basis
while the right-handed Majorana mass matrix is diagonal. Such
reduction to $\epsilon_1$ is possible since the asymmetry at $M_3$
was washed out by the process at $M_2$ and, subsequently, the
asymmetry at $M_2$ was again washed out by the process at $M_1$
\cite{Flanz:1994yx,Covi:1996wh,Buchmuller:1997yu}.

The $\kappa$ in Eq.(\ref{aalepto}) is determined by solving the
full Boltzmann equations. The $\kappa$ can be simply parameterized
in terms of $K$ defined as the ratio of $\Gamma_1$ the decay width
of $\nu_{R1}$ to $H$ the Hubble parameter at temperature $M_1$
\cite{Kolb:vq}, where $K<1$ describes processes out of thermal
equilibrium and $\kappa<1$ describes washout effect,
\bea %
    -\kappa \simeq {0.3 \over {K \left(ln K \right)^{0.6}}},
\eea %
for $10 \lesssim K= \Gamma_1 / H \lesssim 10^6,$ and%
\bea %
    -\kappa \sim
        {1 \over {2 \sqrt{K^2+9}}},
\eea %
for $0 \lesssim K \lesssim 10$
\cite{Kolb:vq}\cite{Pilaftsis:1997jf}\cite{Joshipura:2001ui}. Even
if the decay is much slower than expansion rate of the universe,
there is an unavoidable wash-out effect by $\kappa \sim 0.1$
\cite{Buchmuller:2000as}, the size of which does not vary
significantly as $K$ varies $0$ to $10$. Thus, the lepton
asymmetry in Eq.(\ref{aalepto}) requires a lower limit in CP
asymmetry, to say $\epsilon_1 \gtrsim 10^{-6}$ considering $Y_L
\gtrsim 10^{-9}$.

It has been verified that the decay processes of heavy Majorana
neutrinos can generate a sufficient amount of asymmetry and the
asymmetry could be kept from washout in models with neutrinos
based on the see-saw mechanism
\cite{Luty:un,Joshipura:2001ui,Buchmuller:2000as,Buchmuller:1998zf,Berger:2001np}.
The see-saw mechanism \cite{Gell-Mann:vs} links the low-energy
parameters (experimental observables) to high-energy parameters
(masses of SM singlet neutrinos) through 15 independent parameters
of Yukawa matrix. Whichever the direction of prediction by see-saw
mechanism goes, top-down or bottom-up, the 15 parameters make full
contribution in general to the prediction of any physical
parameters, angles, CP phases, or mass eigenvalues at low energy
or at high energy. So far, on the other hands, the lack in
understanding of Yukawa parameters has restricted consideration of
the low-high link so as to rely on choice of a model. For
instance, there are models which predict low-energy observables in
terms of parameters of leptogenesis
\cite{Branco:2001pq,Rebelo:2002wj,Buchmuller:2002rq,Davidson:2002qv,Berger:2001np}
and/or minimize the degree of freedom in parameter space of
leptogenesis \cite{Frampton:2002qc}\cite{Kaneko:2002yp} at the
present.

The purpose of this paper is to find the structure of matrix of
neutrino Yukawa couplings (Yukawa matrix) which satisfy three
constraints simultaneously: (i) the out-of-equilibrium condition
of leptogenesis, (ii) the upper bound of lepton flavor violating
decay in supersymmetric Standard Model (MSSM), (iii) the
prediction of large mixing angle in atmospheric neutrino
oscillation \cite{Fukuda:2001nk} and CHOOZ bound
\cite{Apollonio:1999ae} of reactor neutrinos through see-saw
mechanism. The structure of Yukawa matrix is understood as
assembly of a matrix of eigenvalues, left transformation and right
transformation. An advantage of the above point of view over mass
structure, to say, in bi-unitary parametrization, compared with
taking a whole mass structure in weak basis, is that the
coexistence of any large mixing angle and hierarchical mass
spectrum can be clearly defined. Another advantage is that the
application of constraints can be systematic because the
constraint (i) is irrelevant to the left transformation while the
constraint (ii) is irrelevant to the right transformation as we
will see. While the constraints (i) and (ii) have separate
dependence on the right mixing and the left mixing, respectively,
the constraint (iii) is connected to both left mixing and right
mixing matrices. A series of phenomenological constraints were
applied to the three parts of Yukawa structure carefully so as to
partition parameter space into an eligible part and a ruled-out
part.

First, the out-of-equilibrium condition of lepton number violating
process constrains the right-handed transformation. In Section II,
an appropriate parametrization for model comparison is introduced
so that the size of CP asymmetry for leptogenesis can be written
in a minimal way. In Section III, the out-of-equilibrium condition
is rephrased in an effective form to fit out purpose and examined
model by model depending on leading contribution which gives rise
to comparison with the bound of the condition. Each model has its
own characteristic range of elements of right transformation and
the expression for the CP asymmetry from the leading contribution
determined in each model.

In Section IV, the strong bound of the branching ratio of $\mu
\rightarrow e \gamma$ in MSSM is briefly reviewed. One can derive
the constraint (ii) on Yukawa couplings focused on left mixing
angles from renormalization group equation(RGE) of scalar mass
terms. As for mixing angles, any pattern of left transformation is
not ruled out by the constraint (ii). When the consideration is
accompanied with the comparison between eigenvalues, a particular
pattern can be ruled out by the constraint. The combination of two
transformation matrices satisfied by (i) and (ii), with the type
of eigenvalues, i.e., nearly degenerate or hierarchical, will be
tested whether they can predict large mixing angle solution in
atmospheric neutrinos and CHOOZ bound in reactor neutrinos through
see-saw mechanism. We have only seven characteristic combinations
which satisfy the three constraints in Tables \ref{table2} and
\ref{table3}. The number of combinations is a consequence after
excluding the cases where a suppressed eigenvalues can be derived
only by fine tuning. There are four such cases shown in Appendix
A.2. From Table \ref{table4} in Appendix A.2, we will see that the
constraint (ii) is redundant if we first apply other two
constraints (i) and (iii).

In Section V, we discuss the direction for the improvement of
models and the possible aspect of application of further
constraints in bi-unitary parametrization. A model will be
examined as an appealing one for leptogenesis.  A summary follows.

\section{Bi-Unitary Transformation of Neutrino Yukawa matrix}
\noindent%
The leptogenesis has attracted attention recently because the
see-saw mechanism is a plausible way for providing the scale for a
lepton number violating process in the early universe. The
observed neutrino mass spectrum in experiment is consistent with a
see-saw model with two scales, $\Lambda_{EW}$ electroweak(EW)
symmetry breaking scale and $\Lambda_{GUT}$ the scale of grand
unified theory(GUT). In the low-energy effective theory, a lepton
number-violating interaction results in light neutrino masses by
non-zero vacuum expectation value of a Higgs scalar, and the mass
matrix is diagonalized by 3 mixing angles, 1 Dirac phase and 2
Majorana phases. As far as the renormalizable gauge invariant
couplings are concerned, the lepton sector will be
\bea %
    \mathcal{L} = -\yuk_E H \overline{\ell} e_R
                    -\yuk_N H \overline{\ell} \nu_R
                    - {1 \over 2} M_R \overline{\nu_R^c} \nu_R + h.c.,
    \label{bblag}
\eea %
in the weak basis. The physical basis of light neutrinos and heavy
neutrinos where the $6 \times 6$ mass matrix is diagonal can be
obtained by a $6 \times 6$ unitary transformation from the weak
basis of left-handed neutrinos in the $\ell$ and right neutrinos
$\nu_R$. This full transformation is reduced to the MNS
\cite{Maki:1962mu} in the low-energy limit, in the basis where
charged lepton mass matrix and right neutrino mass matrix $M_R$
are diagonal.

Considering that the Yukawa matrix is not necessarily diagonal in
the physical basis where masses of Majorana neutrinos are
diagonal, the transformation of Yukawa matrix (neutrino Dirac mass
matrix) needs to be defined in addition to the transformation of
Majorana mass. However, there might exist constraints that can
connect the MNS to left-handed mixing matrix of Yukawa matrix, for
example, arising from a flavor symmetry. Choose a basis where $M_R
= Diag \left( M_1, M_2, M_3 \right)$. Let $\yuk_D = Diag \left(
y_1, y_2, y_3 \right)$, so that a $3 \times 3$ matrix $\yuk_N$ for
three-generation neutrinos can be expressed in use of bi-unitary
transformation as
\begin{eqnarray}
    \yuk_N = \mathbb{L} \yuk_D
    \mathbb{R}^\dagger.
    \label{bbbiunitary}
\end{eqnarray} Pascoli, Petcov, and Rodejohann also discussed this parametrization  in
Ref.\cite{Pascoli:2003uh} in comparison with other
parameterizations to describe leptogenesis. The unitarity of
$\mathbb{L}$ implies that any left-handed mixing angle in Yukawa
matrix does not affect any parameters in leptogenesis, because the
loop contribution appears in terms of $\yuk^\dagger_N \yuk_N =
\unir \yuk_D^2 \unir^\dagger$.
\bea %
    && \left( \yuk^\dagger_N \yuk_N \right)_{11}
    = y_1^2 |R_{11}|^2+ y_2^2 |R_{12}|^2
    + y_3^2 |R_{13}|^2, \label{bbyy1} \label{bbyyk} \\
        && \left(\yuk^\dagger_N \yuk_N \right)_{1k}
        =  y_1^2 R_{11} R_{k1}^*+ y_2^2 R_{12} R_{k2}^*
        + y_3^2R_{13} R_{k3}^*,
\eea%
where $R_{ij}$ is an element of $\mathbb{R}$ and $k = 2$ to $3$.
Expressing the CP asymmetry in Eq.(\ref{aacp}) in terms of the
Yukawa matrices yields,
\bea %
    \epsilon_1 \approx 10^{-1}
        {\rm Im} \left[
            {{\left( \yuk^\dagger_N \yuk_N \right)_{12}}^2
            \over \left( \yuk^\dagger_N \yuk_N \right)_{11} }
            {M_1 \over M_2} +
                {{\left( \yuk^\dagger_N \yuk_N \right)_{13}}^2
                \over \left( \yuk^\dagger_N \yuk_N \right)_{11} }
                {M_1 \over M_3} \right],
    \label{bbepsilon1} \\ \nonumber
\eea %
and all the possible models can then be classified into three
cases depending on the dominant term in Eq.(\ref{bbyy1}). Factors
will be indicated in bold strokes in the following if they are
much smaller than order one. These lead to the following cases.
However, the terms with the bold factors cannot be necessarily
neglected in the leptogenesis analysis. \\
\begin{widetext}
\noindent {\bf Case I}, where $y_3^2 |R_{13}|^2$ is dominant,
\bea %
    {{\left( \yuk^\dagger_N \yuk_N \right)_{1k}}^2
    \over \left( \yuk^\dagger_N \yuk_N \right)_{11}}
    & \approx &
        \left(
            {{y_1^2 R_{11} R_{k1}^*} \over {y_3 |R_{13}|}} +
            {{y_2^2 R_{12} R_{k2}^*} \over {y_3 |R_{13}|}} +
            {{y_3^2 R_{13} R_{k3}^*} \over {y_3 |R_{13}|}}
                \right)^2 \nonumber \\
    & = &
        \left(
            \mathbf{{y_1 |R_{11}|} \over {y_3 |R_{13}|}}
            y_1 R_{k1}^* e^{i\gamma_1}+
                \mathbf{{y_2 |R_{12}|} \over {y_3 |R_{13}|}}
                y_2 R_{k2}^* e^{i\gamma_2} +
                    y_3 R_{k3}^* e^{i\gamma_3}
        \right)^2,
    \label{zzy3dom}
\eea %
where $\gamma_1, \gamma_2,$ and $\gamma_3$ are arguments of
$R_{11}, R_{12},$ and $R_{13}$, respectively.
\newline \newline
\noindent{\bf Case II}, where $y_2^2 |R_{12}|^2$ is dominant,%
\bea %
    {{\left( \yuk^\dagger_N \yuk_N \right)_{1k}}^2
    \over \left( \yuk^\dagger_N \yuk_N \right)_{11}}
    & \approx &
        \left(
            {{y_1^2 R_{11} R_{k1}^*} \over {y_2 |R_{12}|}} +
            {{y_2^2 R_{12} R_{k2}^*} \over {y_2 |R_{12}|}} +
            {{y_3^2 R_{13} R_{k3}^*} \over {y_2 |R_{12}|}}
                \right)^2 \nonumber \\
    & = &
        \left(
            \mathbf{{y_1 |R_{11}|} \over {y_2 |R_{12}|}}
            y_1 R_{k1}^* e^{i\gamma_1}+
                y_2 R_{k2}^* e^{i\gamma_2} +
                    \mathbf{{y_3 |R_{13}|} \over {y_2 |R_{12}|}}
                    y_3 R_{k3}^* e^{i\gamma_3}
                \right)^2,
    \label{zzy2dom}
\eea %
{\bf Case III}, where $y_1^2 |R_{11}|^2$ is dominant,%
\bea %
    {{\left( \yuk^\dagger_N \yuk_N \right)_{1k}}^2
    \over \left( \yuk^\dagger_N \yuk_N \right)_{11}}
    & \approx &
        \left(
            {{y_1^2 R_{11} R_{k1}^*} \over {y_1 |R_{11}|}} +
            {{y_2^2 R_{12} R_{k2}^*} \over {y_1 |R_{11}|}} +
            {{y_3^2 R_{13} R_{k3}^*} \over {y_1 |R_{11}|}}
                \right)^2 \nonumber \\
    & = &
        \left(
            y_1 R_{k1}^* e^{i\gamma_1}+
                \mathbf{{y_2 |R_{12}|} \over {y_1 |R_{11}|}}
                y_2 R_{k2}^* e^{i\gamma_2}+
                    \mathbf{{y_3 |R_{13}|} \over {y_1 |R_{11}|}}
                    y_3 R_{k3}^* e^{i\gamma_3}
                \right)^2,
    \label{zzy1dom}
\eea %
and {\bf Case IV}, where $y_1^2 |R_{11}|^2$, $y_2^2 |R_{12}|^2$,
and $y_3^2 |R_{13}|^2$ are equally dominant,
\bea %
    {{\left( \yuk^\dagger_N \yuk_N \right)_{1k}}^2
    \over \left( \yuk^\dagger_N \yuk_N \right)_{11}}
    & \approx &
        \left(
            {{y_1^2 R_{11} R_{k1}^*} \over {y_1 |R_{11}|}} +
            {{y_2^2 R_{12} R_{k2}^*} \over {y_2 |R_{12}|}} +
            {{y_3^2 R_{13} R_{k3}^*} \over {y_3 |R_{13}|}}
                \right)^2 \nonumber \\
    & = &
        \left(
            y_1 R_{k1}^* e^{i\gamma_1}+
                y_2 R_{k2}^* e^{i\gamma_2}+
                    y_3 R_{k3}^* e^{i\gamma_3}
                \right)^2.
        \label{zznodom}
\eea %
\end{widetext}%
There are also cases with two equally dominant terms in $\left(
\yuk_N^\dagger \yuk_N \right)_{11}$ in Eq.(\ref{bbyy1}).

{\bf Case II*}, where $y_3^2 |R_{13}|^2, y_2^2 |R_{12}|^2 \gg
y_1^2 |R_{11}|^2$,

{\bf Case III*}, where $y_3^2 |R_{13}|^2, y_1^2 |R_{11}|^2 \gg
y_2^2 |R_{12}|^2$,

and {\bf Case IV*}, where $y_1^2 |R_{11}|^2, y_2^2 |R_{12}|^2 \gg
y_3^2 |R_{13}|^2$.

\noindent The corresponding expression to ${\left( \yuk^\dagger_N
\yuk_N \right)_{1k}}^2 / \left( \yuk^\dagger_N \yuk_N
\right)_{11}$ can be found with a single suppressed factor. Such 3
possible cases have similar aspect to the cases \textbf{II, III},
and \textbf{IV} when the out-of-equilibrium condition of lepton
number violating process constrains the magnitude of $\unir$, to
be discussed further in the next section.

The hermitian operator $\yuk^\dagger_N \yuk_N$ can be expressed
with three independent CP phases and six real
parameters such as %
\bea %
    \yuk^\dagger_N \yuk_N =
        \left( \begin{array}{ccccc}
            z_{11} & & z_{12} e^{i \phi_1} & & z_{13} e^{i \phi_2} \\
            z_{12} e^{-i\phi_1} & & z_{22} & & z_{23} e^{i \phi_3}\\
            z_{13} e^{-i\phi_2} & & z_{23} e^{-i\phi_3} & & z_{33} \\
        \end{array} \right),
    \label{ccyy}
\eea %
where the $z_{ij}$'s are real. Two phases, say $\phi_1$ and
$\phi_2$, can be eliminated by a diagonal phase transformation, $
\mathbb{P} \equiv Diag \left(1, exp(-i \phi_1), exp(-i \phi_2)
\right)$.
\bea %
    \mathbb{P}^\dagger \yuk^\dagger_N \yuk_N \mathbb{P} =
        \left( \begin{array}{ccccc}
            z_{11} & & z_{12} & & z_{13} \\
            z_{12} & & z_{22} & & z_{23} e^{i \delta'}\\
            z_{13} & & z_{23} e^{-i\delta'} & & z_{33} \\
        \end{array} \right),
    \label{ccpyyp}
\eea %
where $\delta' \equiv -\phi_1 -\phi_2 +\phi_3$. In general, the
diagonalization of $\mathbb{P} \yuk_N^\dagger \yuk_N
\mathbb{P}^\dagger$ still requires 3 independent phases as
combination of 2 Majorana phases and 1 Dirac phase. One can take
the absolute values of the elements of $\unir$ in
Eqs.(\ref{zzy3dom})-(\ref{zzy1dom}) to find $|{\yuk_N}^\dagger
\yuk_N|_{1k}^2$, regardless of the complexity with $\gamma_1,
\gamma_2$, and $\gamma_3$ in Eqs.(\ref{zzy3dom})-(\ref{zzy1dom})
in terms of elements of $\unir$. For example, the size of CP
asymmetry in Eq.(\ref{bbepsilon1}) can be written as,
\bea %
    \epsilon_1
        & \approx & 10^{-1}
            {{| \yuk^\dagger_N \yuk_N |_{12}}^2
            \over { \yuk^\dagger_N \yuk_N }_{11} }
                \sin 2 \phi_1
                {M_1 \over M_2}
    \nonumber \\
        & + & 10^{-1}
            {{| \yuk^\dagger_N \yuk_N |_{13}}^2
            \over { \yuk^\dagger_N \yuk_N }_{11} }
                \sin 2 \phi_2
                {M_1 \over M_3},
    \label{ccphase}
\eea %
using Eq.(\ref{ccyy}).

\section{Out-of-Equilibrium Condition and Its constraints on
Yukawa mixing angles}

\noindent The decay width of $\nu_{R1}$ by the Yukawa interaction
at tree level is
\bea %
    \Gamma_1 =
        {1 \over {8 \pi}}
        \left(\yuk^\dagger_N\yuk_N
        \right)_{11} M_1.
\eea%
The out of equilibrium condition results when the Hubble parameter
exceeds the decay rate, expressing Hubble parameter in terms of
temperature $T$,
\bea%
    \Gamma_1  <  H = 1.66 g^{1/2}_*{T^2 \over
        M_{pl}}.
\eea %
At temperature $T = M_1$, the condition can be rephrased as,
\bea %
    \left( \yuk^\dagger_N \yuk_N \right)_{11}
             \alt \zeta_1^2,
    \label{ddoffeq}
\eea%
where
\begin{eqnarray}
\zeta_1^2 \equiv 10^2 {M_1  \over M_{pl}}.
\end{eqnarray}
The upper bound $\zeta_1^2$ in Eq.(\ref{ddoffeq}) cannot exceed
$M_1 / M_{GUT}$, when $M_{GUT}$ is chosen in a range $\left(
10^{-4} - 10^{-3} \right) M_{Pl}$ depending on a theoretical
framework and the masses of heavy Majorana neutrinos cannot be
higher than $M_{GUT}$. As long as $M_1$ is hierarchically smaller
than the other masses of heavy neutrinos, as assumed here, the
$\zeta_1^2$ should be considered safely as small as order of $
\lesssim 10^{-2}$.

The condition was examined in detail by Buchmuller and Plumacher
\cite{Buchmuller:1997yu} with the relation of generated $B-L$
asymmetry to the effective mass defined as%
\bea%
    \tilde{m}_1 = \left( \yuk^\dagger_N \yuk_N \right)_{11}
                {v_2^2 \over M_1}.
\eea%
Besides the upper bound which keeps the asymmetry protected from
wash-out effect, there is also a lower bound for $\tilde{m}_1$
since, at high temperature, weak Yukawa couplings cannot produce
enough neutrinos. The eligible $\tilde{m}_1$ capable of generating
sufficient $B-L$ (or equivalently $L$) asymmetry, lies in the
range $m_1 < \tilde{m}_1 < m_3$ due to see-saw mechanism
\cite{Berger:2001np}\cite{Buchmuller:2002rq}, where $m_1$ is the
mass of the lightest neutrino, which is presumed to be non-zero
but smaller than $m_2 \approx \sqrt{\triangle m_\odot^2}$ and
$m_3$ is the mass of the heaviest light neutrino whose magnitude
is about $\sqrt{\triangle m_{atm}^2}$, assuming a hierarchical
light neutrino spectrum.

{\squeezetable
\begin{table*} \caption{Model classification: \textbf{Cases I, II,
III}, and \textbf{IV} are introduced in
Eqs.(\ref{zzy3dom}-\ref{zznodom}). \textbf{Case -a} is for $y_3
\gg y_2 \gg y_1$, \textbf{Case -b} for $y_3 \sim y_2 \gg y_1$, and
\textbf{Case -c} for $y_3 \gg y_2 \sim y_1$. In $|\unir|$,
$\zeta_1$ is defined in Eq.(\ref{ddoffeq}), $\zeta_s$'s in
\textbf{Cases I-b, IV-b} and in \textbf{Case II-b} represent small
values of $|R_{13}|$ and $|R_{12}|$, respectively, while $\zeta_s$
in \textbf{Case III} and \textbf{Case IV-a} represent simply a
small angle. The $s_{12}$ and $s_{23}$ imply that those angles can
be large or small. The \textbf{Case II} and the \textbf{Case -c}
cannot be compatible with each other.}
\begin{ruledtabular}
\begin{tabular}{lccccc}
    & & Models &
    & $ |\unir| \sim $
    & $\epsilon_1$ in terms of
    leading contribution \\
    \hline \\
        \textbf{I-a,-c}
        & & $ \{ \begin{array}{l}  y_3^2|R_{13}|^2 \gg
        y_2^2|R_{12}|^2, y_1^2|R_{11}|^2
        \\ {y_3 \gg y_2 \gg y_1 } $ or $ {y_3 \gg y_2 \sim y_1 } \end{array} $
        & & $\left[
        \begin{array}{ccc}
         1 & s_{12} & <\zeta_1 \\
         s_{12} & 1 & s_{23} \\
         s_{12}s_{23} & s_{23} & 1
        \end{array} \right]$
        & $ 10^{-1} \left(y_3|R_{23}|\right)^2 {M_1 \over M_2}$ \\
            \\
            \textbf{I-b}
            & & $ \{ \begin{array}{l}  y_3^2|R_{13}|^2 \gg
            y_2^2|R_{12}|^2, y_1^2|R_{11}|^2
            \\ {y_3 \sim y_2 \gg y_1 } \end{array} $
            & & $\left[
            \begin{array}{ccc}
            1 & \ll \zeta_s & <\zeta_1 \\
            \ll \zeta_s & 1 & s_{23} \\
            \ll \zeta_s & s_{23} & 1
            \end{array} \right]$
            & $ 10^{-1} \left(y_3|R_{23}|\right)^2 {M_1 \over M_2}$ \\
                \\
                \textbf{II-a}
                & & $ \{ \begin{array}{l}  y_2^2|R_{12}|^2 \gg
                y_3^2|R_{13}|^2, y_1^2|R_{11}|^2
                \\ {y_3 \gg y_2 \gg y_1 } \end{array} $
                & & $\left[
                \begin{array}{ccc}
                1 & s_{12} & <\zeta_1 \\
                s_{12} & 1 & s_{23} \\
                s_{12}s_{23} & s_{23} & 1
                \end{array} \right]$
                & $ 10^{-1} \{ y_2|R_{22}|
                + \mathbf{y_3|R_{13}|\over y_2|R_{12}|}y_3|R_{23}| \}^2
                {M_1 \over M_2}$ \\
            \\
            \textbf{II-b}
            & & $ \{ \begin{array}{l}  y_2^2|R_{12}|^2 \gg
            y_3^2|R_{13}|^2, y_1^2|R_{11}|^2
            \\ {y_3 \sim y_2 \gg y_1 } \end{array} $
            & & $\left[
            \begin{array}{ccc}
            1 & <\zeta_1 & \ll \zeta_s \\
            <\zeta_1 & 1 & s_{23} \\
            <\zeta_1 & s_{23} & 1
            \end{array} \right]$
            & $ 10^{-1} \left( y_2|R_{22}| \right)^2 {M_1 \over M_2} $ \\
        \\
        \textbf{III-a,-b,-c}
        & & $ \{ \begin{array}{l}  y_1^2|R_{11}|^2 \gg
        y_2^2|R_{12}|^2, y_3^2|R_{13}|^2
        \\ {y_3 \gg y_2 \gg y_1 },
        {y_3 \sim y_2 \gg y_1 },
        \\ $ or  $ {y_3 \gg y_2 \gg y_1 }\end{array} $
        & & $\left[
        \begin{array}{ccc}
         1 & \zeta_s & <\zeta_1 \cr
         \zeta_s & 1 & s_{23} \\
         \zeta_s & s_{23} & 1
        \end{array} \right]$
        & $ 10^{-1} \{
        y_1|R_{21}| +
        \mathbf{y_2|R_{12}| \over y_1|R_{11}|}y_2|R_{22}|+
        \mathbf{y_3|R_{13}| \over y_1|R_{11}|}y_3|R_{23}|
        \} ^2
        {M_1 \over M_2} $ \\
                \\
                \textbf{IV-a,-b}
                & & $ \{ \begin{array}{l}  y_3^2|R_{13}|^2 \sim
                y_2^2|R_{12}|^2 \sim y_1^2|R_{11}|^2
                \\ {y_3 \gg y_2 \gg y_1} $ or $ {y_3 \sim y_2 \gg y_1 }
                \end{array} $
                & & $\left[
                \begin{array}{ccc}
                1 & \zeta_s & <\zeta_1 \\
                \zeta_s & 1 & s_{23} \\
                \zeta_s & s_{23} & 1
                \end{array} \right]$
                & $ 10^{-1} \{ y_2|R_{22}|
                + y_3|R_{23}| \}^2
                {M_1 \over M_2}$ \\
            \\
            \textbf{IV-c}
                & & $ \{ \begin{array}{l}  y_3^2|R_{13}|^2 \sim
                y_2^2|R_{12}|^2 \sim y_1^2|R_{11}|^2
                \\ {y_3 \gg y_2 \sim y_1} \end{array} $
                & & $\left[
                \begin{array}{ccc}
                1 & 1 & <\zeta_1 \\
                1 & 1 & s_{23} \\
                s_{23} & s_{23} & 1
                \end{array} \right]$
                & $ 10^{-1} \{ y_2|R_{22}|
                + y_3|R_{23}| \}^2
                {M_1 \over M_2}$ \\ \\
\end{tabular}
\end{ruledtabular}\label{table1}
\end{table*}}

Using Eq.(\ref{bbyy1}), the condition in Eq.(\ref{ddoffeq}) can be
rephrased in
such a convenient way as %
\bea %
    y_1^2 |R_{11}|^2 + y_2^2 |R_{12}|^2 + y_3^2 |R_{13}|^2
    \lesssim \zeta_1^2.
    \label{ddoffeq2}
\eea %
The first significant implication of the above condition is %
\bea %
    |R_{13}| \lesssim \zeta_1,
    \label{ddcondr13}
\eea %
when $y_3$ is of order one. The hierarchy in masses of
right-handed neutrinos is a preliminary for the CP asymmetry in
Eq.(\ref{aacp1}) of leptogenesis, causing the smallness of
$\zeta_1$ and so a small-angle constraint for $ |R_{13}| $. The
next implication is the smallness
of $y_1$,%
\bea %
    y_1 \lesssim \zeta_1,
    \label{ddcondy1}
\eea %
since $|R_{11}|$ is always of order one whether mixing angles are
large or small. So $y_1$ cannot be of the same approximate size as
$y_3$ if $y_3 \sim 1$. The models that satisfy the two conditions
in Eqs.(\ref{ddcondr13})-(\ref{ddcondy1}) can be according to
which term dominates in the real number $(\yuk^\dagger_N
\yuk_N)_{11}$ and how different the eigenvalues $y_1, y_2,$ and
$y_3$ of the Yukawa matrix are. A model with $y_3 \gg y_2 \gg y_1$
is denoted as \textbf{Case -a}, the one with $y_3 \sim y_2 \gg
y_1$ as \textbf{Case -b}, and the one with $y_3 \gg y_2 \sim y_1$
as \textbf{Case -c}. The leading contribution to $\epsilon_1$ in
Table \ref{table1} is presented by the first term in
Eq.(\ref{bbepsilon1}).

In \textbf{Case I} in Eq.(\ref{zzy3dom}), if the eigenvalues are
of normal hierarchy, $y_3 \gg y_2$, the out-of-equilibrium
condition in Eq.(\ref{ddoffeq2}) does not place any constraints on
elements in $\unir$ other than $|R_{13}| < \zeta_1$, while nearly
degenerate eigenvalues, $y_3 \sim y_2$, restrict $|R_{12}|$ to be
small. If the transformation $\unir / \mathbb{J} \equiv \unir
\mathbb{J}^{-1}$ is parameterized as
\begin{widetext}
\bea
    \unir / \mathbb{J} =
    \left(
    \begin{array}{ccc}
        c_{13}c_{12} &  s_{12}c_{13} & s_{13} \\
        -s_{12}c_{23}-s_{23}s_{13}c_{12}e^{-i\delta} &
        c_{23}c_{12}-s_{23}s_{13}s_{12}e^{-i\delta} & s_{23}c_{13}e^{-i\delta} \\
        s_{23}s_{12}e^{i\delta}-s_{13}c_{23}c_{12} &
        -s_{23}c_{12}e^{i\delta}-s_{13}s_{12}c_{23} & c_{23}c_{13} \\
    \end{array}
    \right),
\eea
\end{widetext}
where $\mathbb{J}$ is a diagonal phase transformation, one can
deduce the smallness of $|R_{21}|$ and $|R_{31}|$ from the
smallness of $R_{12}$ due to small $s_{12}$. The possible
structures of $\unir$ for the \textbf{Case I} are listed in Table
\ref{table1}, where $s_{12}$ and $s_{23}$ specified in the table
are allowed to be either small angle or large one, while
$\lambda_s$ indicates that only a small angle is allowed. If
$s_{23}$ is not so small, the leading contribution of the
imaginary part which leads to $\epsilon_1$ consists of mainly
$\left(y_3|R_{23}| \right)^2 $ and $M_1 / M_2$. The model examined
in Ref.\cite{Berger:2001np} is an example of \textbf{Case I-a}.

In \textbf{Case II}, in Eq.(\ref{zzy2dom}), if the eigenvalues are
of normal hierarchy with $y_3 \gg y_2$, the out-of-equilibrium
condition allows $|R_{12}|$ to be much larger than $|R_{13}|$. If
the assumption is made that $y_2$ is not larger than $\zeta_1$,
then $|R_{12}|$ can still be a large angle. In other words,
$|R_{12}|$ can be either small or large, though, it is larger than
$\zeta_1$ in this model. The leading contribution to $\epsilon_1$
depends on
mainly from %
\bea %
y_2|R_{22}| + \mathbf{y_3|R_{13}| \over y_2|R_{12}|}y_3|R_{23}|.
\eea %
Otherwise, the two eigenvalues, $y_2$ and $y_3$ in \textbf{Case
II}, are close to each other. The element $|R_{12}|$ is smaller
than $\zeta_1$ and $|R_{13}|$ is far smaller than $|R_{12}|$. The
term ${| \yuk^\dagger_N \yuk_N |_{12}}^2 / \left({ \yuk^\dagger_N
\yuk_N }\right)_{11}$ reduces to $y_2|R_{22}|$ which is now the
leading contribution to $\epsilon_1$ of order one. The assumption
implies $|R_{12}| \gg y_1/y_2$, so that the \textbf{Case II-c}
with $y_1 \sim y_2$ is ruled out. For the same reason, there can
not be \textbf{Case II*-c}, while \textbf{Case II*-a} with $y_3
\gg y_2$ and \textbf{Case II*-b} with $y_3 \sim y_2$ and small
$|R_{12}|$ are allowed. The allowed ranges in the elements of
$|\unir|$ for the \textbf{Case II*-a} and \textbf{Case II*-b} can
be obtained as the same as in the \textbf{Case II-a} and
\textbf{Case II-b}, respectively.

\textbf{Case III} in Eq.(\ref{zzy1dom}) corresponds to $|R_{12}|
\ll y_1/y_2$, i.e., $|R_{12}|$ is always a small angle. Any part
of the contribution to ${| \yuk^\dagger_N \yuk_N |_{12}}^2 /
\left({ \yuk^\dagger_N \yuk_N }\right)_{11}$ is not yet ruled out
by the assumptions given here. The \textbf{Case III*} also gives
rise to similar constraint on $|R_{12}|$ as \textbf{Case III}
does.

In \textbf{Case IV} in Eq.(\ref{zznodom}), a subcase with $y_3
\sim y_2$ and $|R_{12}|$ is not allowed by out-of-equilibrium
condition because $y_2|R_{12}|$ should not be larger than
$\zeta_1$. This is the case with $|R_{12}| \sim y_1/y_2$. Thus,
$|R_{12}|$ is small for $y_2 \gg y_1$, while $|R_{12}|$ is large
for $y_2 \sim y_1$. The possible structures of $|\unir|$ derived
in \textbf{Case IV*} also is consistent withe those in
\textbf{Case IV}.

The structures of $|\unir|$ in \textbf{Cases II*, III*, IV*}
attribute their correspondences to those in \textbf{Cases II, III,
IV}, respectively, to relative significance in comparison of
$y_1|R_{11}|$ and $y_2|R_{12}|$ when $y_3|R_{13}|$ is fixed. In
other word, once $|R_{13}|$ is bounded, cases can be effectively
classified into a case with $y_1|R_{11}| \ll y_2|R_{12}|$, a case
with $y_2|R_{12}| \ll y_1|R_{11}|$, and a case with $y_1|R_{11}|
\sim y_2|R_{12}|$. Thus, the cases with two equal dominant terms
can remain without further detailed discussion.

In Table \ref{table1}, where the various cases are considered, the
allowed structures of $|\unir|$ under those cases and the CP
asymmetry $\epsilon_1$ in terms of leading contribution after
ruling out the suppressed part under the assumptions are
summarized. In \textbf{Case I} and \textbf{Case IV} with
$|R_{23}|$ of order one , or in \textbf{Case II-b}, $\epsilon_1$
can reach its model independent maximum value,
\begin{equation}
    \epsilon_{1max} \sim 10^{-1}{M_1 \over M_2},
    \label{ccmaxcp}
\end{equation}
where $M_1 < M_2$.

\section{Low-energy observables and their constraints on Yukawa mixing
transformations}

\noindent The current experimental limit on the branching
ratio(\br) of the lepton flavor violating(LFV) decay mode, $\mu
\rightarrow e \gamma$ is
\bea %
    & BR \left( \mu \rightarrow e \gamma \right) < 1.2 \times
    10^{-11}.
    \label{ddmega}
\eea %
In the SM, the branching ratio is $< 10^{-50}$ suppressed far
below the observable bound \cite{Petcov:1976ff}. Another
theoretical framework where one can estimate the size of the LFV
decay rates is the minimal Supersymmetry(SUSY) Standard
Model(MSSM), where the rates for LFV processes can be enhanced due
to large $\tan{\beta}$ \cite{Hisano:1995cp}. The neutrino Yukawa
couplings cause the renormalization group equations (RGE) to
develop flavor changing contribution to soft masses of scalar
leptons, which may originate at a universal value for all kinds
scalar leptons at GUT scale $M_X$. The generated LFV mass terms
for scalar leptons after integrating RGE is
\bea %
    \left(\triangle m^2_{\tilde{L}} \right)_{ij}
        & \approx &
        {-1 \over 8 \pi^2} \left(3m_0^2 + A_0^2 \right)
        \sum_{k} \yuk_{Nik} \log{M_X \over M_k}
                 \yuk_{Nkj}^\dagger \nonumber \\
        & \sim &
        {-1 \over 8 \pi^2} \left(3m_0^2 + A_0^2 \right)
        \log{M_X \over M_3}
            \left( \yuk_N \yuk_N^\dagger \right)_{ij},
\eea %
where $\log{M_X \over M_k}$ for $k=1-3$ can be considered as of
the same order even when $M_k$'s are of hierarchy. The $m_0^2$ and
$A_0^2$ are the universal masses and universal trilinear
couplings, respectively, in soft SUSY-breaking lagrangian
\cite{Hisano:1995cp}. The flavor changing decay in
Eq.(\ref{ddmega}) involves neutralino exchange and chargino
exchange in loop diagrams, and the resulting amplitude is
proportional to $\tan{\beta}$. Therefore the branching ratios can
be expressed in terms of $\yuk_N \yuk_N^\dagger$ as follows;
\bea %
    && BR \left(\ell_i \rightarrow \ell_j \gamma \right) \sim
    \\
    && {\alpha^3 \over {G_F^2 m_s^8}}
            | {-1 \over 8 \pi^2}
            \left( 3m_0^2 + A_0^2 \right)
            \log{M_X \over M_3}
                \left( \yuk_N \yuk^\dagger_N \right)_{ij} |^2
                \tan^2 \beta \nonumber,
\eea %
where $m_s$ is the typical mass of a superparticle, $\alpha$ is
the fine structure constant, and $G_F$ is the Fermi constant
\cite{Hisano:1995cp}\cite{Casas:2001sr}. Recently, LFV in
low-energy has been considered for its relevance with leptogenesis
\cite{Kaneko:2002yp}\cite{Ellis:2002fe}.

Even if a model can predict an enhanced branching ratio near the
range accessible in a near future experiment, the eligible
structure of Yukawa matrix implies there might be a constraint for
severely suppressed mixing angles in left-handed transformation of
Yukawa matrix at low energy. The enhancement mechanism of
branching ratios is possible not only by raising $tan{\beta}$ but
also by increasing mixing angles. Hisano et al
\cite{Hisano:1995cp}. analyzed the CKM matrix of quark mixing.
With small mixing angles as in CKM, the branching ratios of
$\ell_i \rightarrow \ell_j \gamma$ processes for large
$\tan{\beta}$ can reach close to the current experimental bounds
in Eq.(\ref{ddmega}). Considering the strong upper bound in
current experiment limit for the decay mode $\mu \rightarrow e
\gamma$, the relevant constraint on the Yukawa couplings is, in
terms of the elements of $\unil$ defined in
Eq.(\ref{bbbiunitary}),
\begin{eqnarray}
    & & \left( \yuk_N \yuk_N^\dagger \right)_{21} =
    \nonumber \\
    & & y_1^2 L_{21}L_{11}^* + y_2^2 L_{22}L_{12}^*
    + y_3^2 L_{23}L_{13}^* \ll 1,
    \label{ddleftcondition}
\end{eqnarray}
which can be satisfied only if $y_2$ or $|L_{12}|$ is small and
$|L_{23}|$ or $|L_{13}|$ is small.

In the rest of this section, I combine the constraint on $\unir$
from the out-of-equilibrium condition and the constraint on
$\unil$ from the upper bound of the branching ratio of $\mu
\rightarrow e \gamma$. The question is whether the result of those
combinations can be accommodated with the experimentally observed
large mixing angles of light neutrinos when implemented via
see-saw mechanism. Light neutrino masses can be obtained through
the see-saw mechanism where Dirac masses are defined in terms of
the non-zero vacuum expectation value $v_2$ of a light Higgs in
MSSM with two Higgs doublets,
\bea %
     m_\nu &=&
            -v_2^2 \yuk_N M_R^{-1} \yuk_N^T, \nonumber \\
           &=&
            -v_2^2 \unil \yuk_D \unir^\dagger M_R^{-1}
            \unir^* \yuk_D^T \unil^T.
    \label{ddseesaw}
\eea %
In this basis the mass matrix of charged leptons is diagonal. When
the transformations $\unil$ and $\unir$ are trivial, i.e., equal
to the identity, the see-saw mechanism yields the light neutrino
mass matrix,
\bea %
    m_\nu = - v_2^2
    Diag \left({y_1^2 \over M_1},{y_2^2 \over M_2},
                {y_3^2 \over M_3} \right),
\eea %
from Eq.(\ref{ddseesaw}). It will be interesting to watch how
neutrino mass matrix in Eq.(\ref{ddseesaw}) can change as the type
of transformation matrix $\unir$, and subsequently $\unil$,
switches from one to another general form. There are 4 types of
$\unir$'s, which have been given in Table \ref{table1}, and 4
types of $\unil$'s are as follows:
\begin{eqnarray}
    && |\unir_1| \sim
            \left( \begin{array}{ccc}
                1 & \rho & \rho \\
                \rho & 1 & 1 \\
                \rho & 1 & 1 \end{array} \right),\hspace{5pt}
        |\unil_1| \sim
            \left( \begin{array}{ccc}
                1 & \lambda & \lambda \\
                \lambda & 1 & 1 \\
                \lambda & 1 & 1 \end{array} \right),
                    \nonumber
\end{eqnarray}
\begin{eqnarray}
    && |\unir_2| \sim
            \left( \begin{array}{ccc}
            1 & 1 & \rho \\
            1 & 1 & \rho \\
            \rho & \rho & 1 \end{array} \right),\hspace{5pt}
        |\unil_2| \sim
            \left( \begin{array}{ccc}
            1 & 1 & \lambda \\
            1 & 1 & \lambda \\
            \lambda & \lambda & 1 \end{array} \right),
               \label{ddunir123} \label{ddunil1234}
\end{eqnarray}
\begin{eqnarray}
    && |\unir_3| \sim
            \left( \begin{array}{ccc}
                1 & 1 & \rho \\
                1 & 1 & 1 \\
                1 & 1 & 1 \end{array} \right),\hspace{5pt}
        |\unil_3| \sim
        \left( \begin{array}{ccc}
                1 & 1 & \lambda \\
                1 & 1 & 1 \\
                1 & 1 & 1 \end{array} \right),
                    \nonumber
\end{eqnarray}
\begin{eqnarray}
    && |\unir_4| \sim
            \left( \begin{array}{ccc}
                1 & \rho & \rho \\
                \rho & 1 & \rho \\
                \rho & \rho & 1 \end{array} \right),\hspace{5pt}
        |\unil_4| \sim
          \left( \begin{array}{ccc}
                1 & \lambda & \lambda \\
                \lambda & 1 & \lambda \\
                \lambda & \lambda & 1 \end{array} \right),
        \nonumber
\end{eqnarray}
which satisfy the out-of-equilibrium condition in
Eq.(\ref{ddoffeq}). The most simplified notation is used, which is
that diagonal elements and large mixing elements are denoted by 1,
whereas small mixing elements are denoted by arbitrary small
values by $\rho \ll 1$ and $\lambda \ll 1$ in $\unir$ and $\unil$,
respectively. That is, there are only two kinds of elements in a
transformation matrix: order one or far less than order one. The
more detailed $\unir$'s have been sorted in Table \ref{table1} by
considering out-of-equilibrium condition. Later, in case the
leading order in an entry of light neutrino mass matrix originates
from small angles in transformation matrices, the light neutrino
mass matrices in terms of the small angles, considering the actual
entries as expansion of small angles, are shown for models in
Appendix A.1.

{\squeezetable
\begin{table*}
\caption{List of possible structures of matrix $m_\nu / v_2^2$
which can be produced by see-saw mechanism in Eq.(\ref{ddseesaw})
using $\unir$'s and $\unil$'s in Eq.(\ref{ddunil1234}), where
$m_\nu$ is a matrix mass of light neutrinos and $v_2$ is a vacuum
expectation value in two-Higgs-doublet models. $M_{123}$ and
$M_{12}$ are defined in Eq.(\ref{ddm123}) and $y_{123}$ and
$y_{12}$ are defined in Eq.(\ref{ddy123}). Please see Appendix A.1
for the entries marked by $\ast$. \vspace{4pt}}
\begin{ruledtabular}
\begin{tabular}{lcc} \\
    & $\unil_1$
    & $\unil_2$ \\
    \\ \hline \\
        $\unir_1$ &
            $ \left[ \begin{array}{ccc}
              y_1^2 M_1^{-1} & \ast & \ast \\
              \ast & y_{23}^2 M_{23}^{-1} & y_{23}^2 M_{23}^{-1}
              \\
              \ast & y_{23}^2 M_{23}^{-1} & y_{23}^2 M_{23}^{-1}
              \end{array} \right]
              $\footnote{\textbf{Model A}: ($\unir_1,\unil_1$), $y_2 \sim y_3$} &
            $ \left[ \begin{array}{ccc}
            y_1^2 M_1^{-1} + y_2^2 M_{23}^{-1} &
            y_1^2 M_1^{-1} + y_2^2 M_{23}^{-1} &
            y_2 y_3 M_{23}^{-1} \\
            y_1^2 M_1^{-1} + y_2^2 M_{23}^{-1} &
            y_1^2 M_1^{-1} + y_2^2 M_{23}^{-1} &
            y_2 y_3 M_{23}^{-1} \\
            y_2 y_3 M_{23}^{-1} & y_2 y_3 M_{23}^{-1} &
            y_3^2 M_{23}^{-1}
            \end{array} \right] $ \\ \\
    \hline \\
        $\unir_2$ &
            $ \left[ \begin{array}{ccc}
              y_1^2 M_{12}^{-1} & y_1 y_2 M_{12}^{-1} &
              y_1 y_2 M_{12}^{-1} \\
              y_1 y_2 M_{12}^{-1} &
              y_2^2 M_{12}^{-1}+y_3^2 M_3^{-1} &
              y_2^2 M_{12}^{-1}+y_3^2 M_3^{-1} \\
              y_1 y_2 M_{12}^{-1} &
              y_2^2 M_{12}^{-1}+y_3^2 M_3^{-1} &
              y_2^2 M_{12}^{-1}+y_3^2 M_3^{-1}
              \end{array} \right] $ \footnote{\textbf{Model B}: ($\unir_2,\unil_1$), $y_2 \ll y_3$} &
              $ \left[ \begin{array}{ccc}
            y_{12}^2 M_{12}^{-1} & y_{12}^2 M_{12}^{-1} & \ast \\
            y_{12}^2 M_{12}^{-1} & y_{12}^2 M_{12}^{-1} & \ast \\
            \ast & \ast & y_3^2 M_3^{-1}
            \end{array} \right] $ \\ \\
    \hline \\
        $\unir_3$ &
            $ \left[ \begin{array}{ccc}
              y_1^2 M_{123}^{-1} &
              y_1 y_2 M_{123}^{-1} + y_1 y_3 M_{23}^{-1} &
              y_1 y_2 M_{123}^{-1} + y_1 y_3 M_{23}^{-1} \\
                  y_1 y_2 M_{123}^{-1} + y_1 y_3 M_{23}^{-1} &
                  y_2^2 M_{123}^{-1} + y_{23} y_3 M_{23}^{-1} &
                  y_2^2 M_{123}^{-1} + y_{23} y_3 M_{23}^{-1} \\
              y_1 y_2 M_{123}^{-1} + y_1 y_3 M_{23}^{-1} &
              y_2^2 M_{123}^{-1} + y_{23} y_3 M_{23}^{-1} &
              y_2^2 M_{123}^{-1} + y_{23} y_3 M_{23}^{-1}
              \end{array} \right] $ &
              $ \left[ \begin{array}{ccc}
            y_{12}^2 M_{123}^{-1} &
            y_{12}^2 M_{123}^{-1} &
            y_{12} y_3 M_{23}^{-1} \\
                y_{12}^2 M_{123}^{-1} &
                y_{12}^2 M_{123}^{-1} &
                y_{12} y_3 M_{23}^{-1} \\
            y_{12} y_3 M_{23}^{-1} &
            y_{12} y_3 M_{23}^{-1} &
            y_3^2 M_{23}^{-1}
            \end{array} \right] $ \\
    \\ \hline \\
        $\unir_4$ &
            $ \left[ \begin{array}{ccc}
              y_1^2 M_1^{-1} & \ast & \ast \\
              \ast &
              y_2^2 M_2^{-1}+y_3^2 M_3^{-1} &
              y_2^2 M_2^{-1}+y_3^2 M_3^{-1} \\
              \ast &
              y_2^2 M_2^{-1}+y_3^2 M_3^{-1} &
              y_2^2 M_2^{-1}+y_3^2 M_3^{-1}
              \end{array} \right] $
              \footnote{\textbf{Model C}: ($\unir_4,\unil_1$), $y_2 \sim y_3$
                        \hspace{2pt}
                        \textbf{Model D}: ($\unir_4,\unil_1$), $y_2 \ll y_3$} &
              $ \left[ \begin{array}{ccc}
            y_1^2 M_1^{-1} + y_2^2 M_2^{-1} &
            y_1^2 M_1^{-1} + y_2^2 M_2^{-1} &
            \ast \\
            y_1^2 M_1^{-1} + y_2^2 M_2^{-1} &
            y_1^2 M_1^{-1} + y_2^2 M_2^{-1} &
            \ast \\
            \ast & \ast &
            y_3^2 M_3^{-1}
            \end{array} \right] $ \\
        \label{table2}
\end{tabular}
\end{ruledtabular}
\vspace{1pt}
\end{table*}}
{\squeezetable
\begin{table*}
\caption{The matrix $m_\nu / v_2^2$ which can be produced by
see-saw mechanism in Eq.(\ref{ddseesaw}) using $\unir$'s and
$\unil$'s in Eq.(\ref{ddunil1234}). The $i$ in the matrix from
$\unir_4$ and $\unil_3$ runs 1 to 3. Please see Appendix A.1 for
the entries marked by $\ast$. \vspace{4pt}}
\begin{ruledtabular}
\begin{tabular}{lcc}\\
    & $\unil_3$
    & $\unil_4$ \\
    \\ \hline \\
        $\unir_1$ &
            $ \left[ \begin{array}{ccc}
              y_1^2 M_1^{-1} + y_2^2 M_{23}^{-1} &
              y_1^2 M_1^{-1} + y_2 y_{23} M_{23}^{-1} &
              y_1^2 M_1^{-1} + y_2 y_{23} M_{23}^{-1} \\
              y_1^2 M_1^{-1} + y_2 y_{23} M_{23}^{-1} &
              y_1^2 M_1^{-1} + y_{23}^2 M_{23}^{-1} &
              y_1^2 M_1^{-1} + y_{23}^2 M_{23}^{-1} \\
              y_1^2 M_1^{-1} + y_2 y_{23} M_{23}^{-1} &
              y_1^2 M_1^{-1} + y_{23}^2 M_{23}^{-1} &
              y_1^2 M_1^{-1} + y_{23}^2 M_{23}^{-1}
              \end{array} \right] $ &
            $ \left[ \begin{array}{ccc}
            y_1^2 M_1^{-1} & \ast & \ast \\
            \ast & y_2^2 M_{23}^{-1} & y_2 y_3 M_{23}^{-1} \\
            \ast & y_2 y_3 M_{23}^{-1} & y_3^2 M_{23}^{-1}
            \end{array} \right] $
            \footnote{\textbf{Model E}: ($\unir_1,\unil_4$), $y_2 \sim y_3$} \\ \\
    \hline \\
        $\unir_2$ &
            $ \left[ \begin{array}{ccc}
              y_{12}^2 M_{12}^{-1} & y_{12}^2 M_{12}^{-1} &
              y_{12}^2 M_{12}^{-1} \\
              y_{12}^2 M_{12}^{-1} &
              y_{12}^2 M_{12}^{-1} + y_3^2 M_3^{-1} &
              y_{12}^2 M_{12}^{-1} + y_3^2 M_3^{-1} \\
              y_{12}^2 M_{12}^{-1} &
              y_{12}^2 M_{12}^{-1} + y_3^2 M_3^{-1} &
              y_{12}^2 M_{12}^{-1} + y_3^2 M_3^{-1}
              \end{array} \right] $
              \footnote{\textbf{Model F}: ($\unir_2,\unil_3$), $y_2 \ll y_3$} &
              $ \left[ \begin{array}{ccc}
            y_1^2 M_{12}^{-1} & y_1 y_2 M_{12}^{-1} & \ast \\
            y_1 y_2 M_{12}^{-1} & y_2^2 M_{12}^{-1} & \ast \\
            \ast & \ast & y_3^2 M_3^{-1}
            \end{array} \right] $ \\ \\
    \hline \\
        $\unir_3$ &
            $ \left[ \begin{array}{ccc}
              y_{12}^2 M_{123}^{-1} &
              y_{12}^2 M_{123}^{-1} + y_{12} y_3 M_{23}^{-1} &
              y_{12}^2 M_{123}^{-1} + y_{12} y_3 M_{23}^{-1} \\
                  y_{12}^2 M_{123}^{-1} + y_{12} y_3 M_{23}^{-1} &
                  y_{12}^2 M_1^{-1} + y_{123}^2 M_{23}^{-1} &
                  y_{12}^2 M_1^{-1} + y_{123}^2 M_{23}^{-1} \\
              y_{12}^2 M_{123}^{-1} + y_{12} y_3 M_{23}^{-1} &
              y_{12}^2 M_1^{-1} + y_{123}^2 M_{23}^{-1} &
              y_{12}^2 M_1^{-1} + y_{123}^2 M_{23}^{-1}
              \end{array} \right] $ &
              $ \left[ \begin{array}{ccc}
              y_1^2 M_{123}^{-1} &
              y_1 y_2 M_{123}^{-1} &
              y_1 y_3 M_{23}^{-1} \\
                  y_1 y_2 M_{123}^{-1} &
                  y_2^2 M_{123}^{-1} &
                  y_2 y_3 M_{23}^{-1} \\
              y_1 y_3 M_{23}^{-1} &
              y_2 y_3 M_{23}^{-1} &
              y_3^2 M_{23}^{-1}
              \end{array} \right] $ \\
 \\ \hline \\
        $\unir_4$ &
            $ \left[ \begin{array}{ccc}
              y_1^2 M_1^{-1} + y_2^2 M_2^{-1} &
              y_1^2 M_1^{-1} + y_2^2 M_2^{-1}&
              y_1^2 M_1^{-1} + y_2^2 M_2^{-1} \\
              y_1^2 M_1^{-1} + y_2^2 M_2^{-1} &
              y_i^2 M_i^{-1} &
              y_i^2 M_i^{-1} \\
              y_1^2 M_1^{-1} + y_2^2 M_2^{-1} &
              y_i^2 M_i^{-1} &
              y_i^2 M_i^{-1}
              \end{array} \right] $
              \footnote{\textbf{Model G}: ($\unir_4,\unil_3$), $y_2 \ll y_3$}  &
            $ \left[ \begin{array}{ccc}
            y_1^2 M_1^{-1} & \ast & \ast \\
            \ast & y_2^2 M_2^{-1} & \ast \\
            \ast & \ast & y_3^2 M_3^{-1}
            \end{array} \right] $ \\
    \label{table3}
\end{tabular}
\end{ruledtabular} \vspace{1pt}
\end{table*}}

{\squeezetable
\begin{table*}
\caption{List of phenomenologically viable models of Yukawa
matrices and the light neutrino mass matrix divided by $v^2$
derived through see-saw mechanism. The eigenvalues in
\textbf{Models A, C,} and \textbf{E} are of $y_2 \ll y_3 \sim 1$,
while those in \textbf{Models B, D, F}, and \textbf{G} are of $y_1
\ll y_2 \sim y_3 \sim 1$. The $y_2$ and $y_3$ specified in $m_\nu
/ v^2$ remain for a later use. \vspace{4pt}}
\begin{ruledtabular}
\begin{tabular}{lcccl} \\
    \textbf{Model} & $|\unil|$ & $\yuk_D$ & $|\unir^\dagger|$ &
    \hspace{10pt} $m_\nu / v^2 \sim \unil \yuk_D \unir^\dagger M_R^{-1}
            \unir^* \yuk_D^T \unil^T $ \\ \\ \hline \\
        \textbf{A} &
        $\left[ \begin{array}{ccc}
               1 & \lambda & \lambda \\
               \lambda & 1 & 1 \\
               \lambda & 1 & 1 \end{array} \right]$ &
        $\left[ \begin{array}{ccc}
               y_1 & &  \\
                   & 1 &  \\
                   &   & 1 \end{array} \right]$ &
        $\left[ \begin{array}{ccc}
                1 & \rho & \rho \\
                \rho & 1 & 1 \\
                \rho & 1 & 1 \end{array} \right]$ &
        $ \left[ \begin{array}{ccc}
              y_1^2 M_1^{-1} & \surd & \surd \\
                ( y_1^2 M_1^{-1} + y_{23}^2 M_{23}^{-1})\lambda
                + y_1 y_{23}M_{123}^{-1}\rho
                & y_{23}^2 M_{23}^{-1} & \surd  \\
                    ( y_1^2 M_1^{-1} + y_{23}^2 M_{23}^{-1})\lambda
                    + y_1 y_{23}M_{123}^{-1}\rho
                    & y_{23}^2 M_{23}^{-1} & y_{23}^2 M_{23}^{-1}
              \end{array} \right] $ \\ \\ \hline \\
        \textbf{B} &
        $\left[ \begin{array}{ccc}
               1 & \lambda & \lambda \\
               \lambda & 1 & 1 \\
               \lambda & 1 & 1 \end{array} \right]$ &
        $\left[ \begin{array}{ccc}
               y_1 & &  \\
                   & y_2 &  \\
                   &   & 1 \end{array} \right]$ &
        $\left[ \begin{array}{ccc}
                1 & 1 & \rho \\
                1 & 1 & \rho \\
                \rho & \rho & 1 \end{array} \right]$ &
        $ \left[ \begin{array}{ccc}
              y_1^2 M_{12}^{-1} & \surd &
              \surd \\
              y_1 y_2 M_{12}^{-1} &
              y_2^2 M_{12}^{-1}+y_3^2 M_3^{-1} &
              \surd \\
              y_1 y_2 M_{12}^{-1} &
              y_2^2 M_{12}^{-1}+y_3^2 M_3^{-1} &
              y_2^2 M_{12}^{-1}+y_3^2 M_3^{-1}
              \end{array} \right] $ \\ \\ \hline \\
        \textbf{C} &
        $\left[ \begin{array}{ccc}
               1 & \lambda & \lambda \\
               \lambda & 1 & 1 \\
               \lambda & 1 & 1 \end{array} \right]$ &
        $\left[ \begin{array}{ccc}
               y_1 & &  \\
                   & 1 &  \\
                   &   & 1 \end{array} \right]$ &
        $\left[ \begin{array}{ccc}
                1 & \rho & \rho \\
                \rho & 1 & \rho \\
                \rho & \rho & 1 \end{array} \right]$ &
        $ \left[ \begin{array}{ccc}
              y_1^2 M_1^{-1} & \surd & \surd \\
              y_i^2 M_i^{-1} \lambda + y_1 \left(y_2 + y_3\right)M_{13}^{-1}\rho &
              y_2^2 M_2^{-1}+y_3^2 M_3^{-1} &
              \surd \\
              y_i^2 M_i^{-1} \lambda + y_1 \left(y_2 + y_3\right)M_{13}^{-1}\rho &
              y_2^2 M_2^{-1}+y_3^2 M_3^{-1} &
              y_2^2 M_2^{-1}+y_3^2 M_3^{-1}
              \end{array} \right] $ \\ \\ \hline \\
        \textbf{D} &
        $\left[ \begin{array}{ccc}
               1 & \lambda & \lambda \\
               \lambda & 1 & 1 \\
               \lambda & 1 & 1 \end{array} \right]$ &
        $\left[ \begin{array}{ccc}
               y_1 & &  \\
                   & y_2 &  \\
                   &   & 1 \end{array} \right]$ &
        $\left[ \begin{array}{ccc}
                1 & \rho & \rho \\
                \rho & 1 & 1 \\
                \rho & 1 & 1 \end{array} \right]$ &
        $ \left[ \begin{array}{ccc}
              y_1^2 M_1^{-1} & \surd & \surd \\
              y_i^2 M_i^{-1} \lambda + y_1 \left(y_2 + y_3\right)M_{13}^{-1}\rho &
              y_2^2 M_2^{-1}+y_3^2 M_3^{-1} &
              \surd \\
              y_i^2 M_i^{-1} \lambda + y_1 \left(y_2 + y_3\right)M_{13}^{-1}\rho &
              y_2^2 M_2^{-1}+y_3^2 M_3^{-1} &
              y_2^2 M_2^{-1}+y_3^2 M_3^{-1}
              \end{array} \right] $ \\ \\ \hline \\
        \textbf{E} &
        $\left[ \begin{array}{ccc}
               1 & \lambda & \lambda \\
               \lambda & 1 & \lambda \\
               \lambda & \lambda & 1 \end{array} \right]$ &
        $\left[ \begin{array}{ccc}
               y_1 & &  \\
                   & 1 &  \\
                   &   & 1 \end{array} \right]$ &
        $\left[ \begin{array}{ccc}
                1 & \rho & \rho \\
                \rho & 1 & 1 \\
                \rho & 1 & 1 \end{array} \right]$ &
        $ \left[ \begin{array}{ccc}
            y_1^2 M_1^{-1} & \surd & \surd \\
            (y_1^2 M_1^{-1} + y_2 y_{23} M_{23}^{-1})\lambda
            + y_1 y_2 M_{123}^{-1}\rho &
            y_2^2 M_{23}^{-1} & \surd \\
                (y_1^2 M_1^{-1} + y_3 y_{23} M_{23}^{-1})\lambda
                + y_1 y_3 M_{123}^{-1}\rho &
                y_2 y_3 M_{23}^{-1} & y_3^2 M_{23}^{-1}
            \end{array} \right] $ \\ \\ \hline \\
        \textbf{F} &
        $\left[ \begin{array}{ccc}
               1 & 1 & \lambda \\
               1 & 1 & 1 \\
               1 & 1 & 1 \end{array} \right]$ &
        $\left[ \begin{array}{ccc}
               y_1 & &  \\
                   & y_2 &  \\
                   &   & 1 \end{array} \right]$ &
        $\left[ \begin{array}{ccc}
                1 & 1 & \rho \\
                1 & 1 & \rho \\
                \rho & \rho & 1 \end{array} \right]$ &
        $ \left[ \begin{array}{ccc}
              y_{12}^2 M_{12}^{-1} & \surd &
              \surd \\
              y_{12}^2 M_{12}^{-1} &
              y_{12}^2 M_{12}^{-1} + y_3^2 M_3^{-1} &
              \surd \\
              y_{12}^2 M_{12}^{-1} &
              y_{12}^2 M_{12}^{-1} + y_3^2 M_3^{-1} &
              y_{12}^2 M_{12}^{-1} + y_3^2 M_3^{-1}
              \end{array} \right] $ \\ \\ \hline \\
        \textbf{G} &
        $\left[ \begin{array}{ccc}
               1 & 1 & \lambda \\
               1 & 1 & 1 \\
               1 & 1 & 1 \end{array} \right]$ &
        $\left[ \begin{array}{ccc}
               y_1 & &  \\
                   & y_2 &  \\
                   &   & 1 \end{array} \right]$ &
        $\left[ \begin{array}{ccc}
                1 & \rho & \rho \\
                \rho & 1 & \rho \\
                \rho & \rho & 1 \end{array} \right]$ &
        $ \left[ \begin{array}{ccc}
              y_1^2 M_1^{-1} + y_2^2 M_2^{-1} &
              \surd &
              \surd \\
              y_1^2 M_1^{-1} + y_2^2 M_2^{-1} &
              y_i^2 M_i^{-1} &
              \surd \\
              y_1^2 M_1^{-1} + y_2^2 M_2^{-1} &
              y_i^2 M_i^{-1} &
              y_i^2 M_i^{-1}
              \end{array} \right] $ \\
    \label{tablesummary}
\end{tabular}
\end{ruledtabular} \vspace{1pt}
\end{table*}}

Each structure of $|\unil_1|-|\unil_4|$ is similar to each of
$|\unir_1|-|\unir_4|$, as shown in Eq.(\ref{ddunir123}), which
satisfy the condition in Eq.(\ref{ddleftcondition}) which is the
same form as Eq.(\ref{ddoffeq2}) for the $\unir$ matrix. For the
suppressed $(\yuk_N \yuk_N^\dagger)_{21}$, it is necessary that
$L_{12}$ and $L_{13}L_{23}$ be small for $y_2 \sim y_3$, while the
smallness of $L_{12}$ or $L_{23}$ is optional for $y_2 \ll y_3$
and $L_{13}$ small. The possibility of large $|L_{13}|$ is
excluded by the CHOOZ bound \cite{Apollonio:1999ae} on the MNS
neutrino mixing matrix.

In Tables \ref{table2} and \ref{table3}, all the possible
structures of light neutrino mass matrix, which can be produced by
Eq.(\ref{ddseesaw}) using $\unir$'s and $\unil$'s in
Eq.(\ref{ddunil1234}), are listed. Due to $M_1 \ll M_2, M_3$
assumed for Eq.(\ref{aacp1}), the following approximation can be
used in the tables.
\begin{eqnarray}
    && M_{123}^{-1} \sim M_{12}^{-1} \sim M_1^{-1}, \nonumber \\
    && M_{123}^{-1} \gg M_{23}^{-1},
    \hspace{5pt} M_{123}^{-1} \gg M_3^{-1},
\end{eqnarray}
where
\begin{eqnarray}
    && M_{123}^{-1} \equiv M_{1}^{-1} + M_{2}^{-1} + M_3^{-1}, \nonumber \\
    && M_{12}^{-1} \equiv M_{1}^{-1} + M_{2}^{-1},
    \label{ddm123}
\end{eqnarray}
etc., and
\begin{eqnarray}
    && y_{123} = y_1 + y_2 + y_3, \nonumber \\
    && y_{12} = y_1 + y_2, \hspace{5pt}
    y_{23} = y_2 + y_3.
    \label{ddy123}
\end{eqnarray}
The comparison between $M_2$ and $M_3$ is not yet fixed. Either
possibility, $M_2 \ll M_3$ or $M_2 \sim M_3$, is still considered.

The structures listed in Tables \ref{table2} and \ref{table3} are
investigated whether they can predict two large mixing angles and
one small mixing angle in light neutrinos. As for the large mixing
angle in atmospheric neutrino oscillation, the ratio of two
elements $m_{\nu 23} /m_{\nu 33}$ must not be less than order one.
On the other hand, for the ratio $m_{\nu 12} /m_{\nu 22}$ to be
not less than order one is a sufficient condition for the large
mixing angle in solar neutrino oscillation, because there is still
a room for a kind of manipulation including fine tuning so as to
make the mixing angle large even without the naive ratio $m_{\nu
12} /m_{\nu 22}$. The application of the constraints from light
neutrino mass spectrum is described more in Appendix A.2. When
$y_2 \sim y_3$, the combination of $\unir$ and $\unil$ which help
us predict large mixing angle for atmospheric neutrinos is
$\left(\unir_1,\unil_1\right)$ for \textbf{Model A},
$\left(\unir_4,\unil_1\right)$ for \textbf{Model C}, or
$\left(\unir_1,\unil_4\right)$ for \textbf{Model E} where
$\unir_i$ and $\unil_i$ with $i = 1 - 4$ are defined in
Eqs.(\ref{ddunir123}). The solutions $\unir_2$ or $\unir_3$ are
not included since the out-of-equilibrium condition with $y_2 \sim
y_3$ ruled out a large angle at $R_{12}$, as shown in Table
\ref{table1}. When $y_2 \ll y_3$, $\left(\unir_2,\unil_1\right)$
for \textbf{Model B}, $\left(\unir_4,\unil_1\right)$ for
\textbf{Model D}, $\left(\unir_2,\unil_3\right)$ for \textbf{Model
F}, $\left(\unir_4,\unil_3\right)$ for \textbf{Model G} can
predict the large mixing angle for atmospheric neutrinos and CHOOZ
bound of reactor neutrinos. See Appendix.

Thus, the structures of Yukawa matrices that satisfy three
constraints:\begin{itemize}
    \item[(i)] the out-of-equilibrium condition of lepton number
    violating processes,
    \item[(ii)] the upper bound of the branching ratio of the rare
    decay $\mu \rightarrow e \gamma$,
    \item[(iii)] the large mixing angle in atmospheric neutrino
    oscillation,
\end{itemize}
must be one of the 7 combinations, \textbf{Models A} to \textbf{G}
in Tables \ref{table2} and \ref{table3}, which establish the
relative dominance of $y_2$ and $y_3$, and the approximate
magnitudes of transformation matrices, $|\unir|$ and $|\unil|$. In
Table \ref{tablesummary} is the list of models of Yukawa matrices
and the light neutrino mass matrix divided by $v^2$. Models are
specified in terms of two unitary transformations and the relative
dominance in eigenvalues as summary of Tables \ref{table2} and
\ref{table3}. The notations are introduced in
Eqs.(\ref{ddunir123}),(\ref{ddm123}),(\ref{ddy123}), and
(\ref{zzmodela})-(\ref{zzmodele}). The entry marked with $\surd$
represents the corresponding symmetric elements. There are two
types of spectrum of eigenvalues, $y_1 \ll y_2, \hspace{4pt} y_1
\sim y_2 \ll y_3 \sim 1$ and $y_1 \ll y_2 \sim y_3 \sim 1$ , which
correspond to \textbf{Case -a,-c} and \textbf{Case -b} introduced
for Table \ref{table1}, respectively.

In the Appendix, Table \ref{table4} describes in detail which
constraints keep certain entries in Tables \ref{table2} and
\ref{table3} from being eligible models. The Table \ref{table4} in
Appendix A.2 shows obviously that the cases prohibited by the
constraint (ii) are already prohibited by the constraint (iii).
The constraint (ii) turns out as a redundant one in checking the
eligibility of Yukawa matrix to known leptonic phenomenology. In
other words, the structure of Yukawa matrix which can derive the
light neutrino mass matrix through see-saw mechanism always appear
to give rise to $Br(\mu \rightarrow e \gamma)$ below the present
experimental limit in SUSY theories with large $\tan\beta$ and
universal slepton mass at GUT scale.  It is possible to pull out a
constraint to narrow down the estimation of heavy neutrino masses
from LFV through the see-saw mechanism.

\section{Models and a link between leptogenesis and low-energy neutrinos}

\subsection{Application of bi-unitary parametrization}
\noindent The light neutrino mass matrix defined through the
seesaw mechanism in Eq.(\ref{ddseesaw}) is diagonalized as
\bea %
    \mns m_L \mns^T =
    -v_2^2 \unil \yuk_D \unir^\dagger M_R^{-1}
            \unir^* \yuk_D^T \unil^T,
    \label{eeseesaw}
\eea %
where $\mns$ is MNS light neutrino transformation matrix and $m_L$
is a diagonal matrix with mass eigenvalues $m_i$. The equation of
$3 \times 3$ symmetric complex matrices in Eq.(\ref{eeseesaw})
consists of 12 equations of parameters, while its right hand side
consists of 18 unknowns: 3 angles and 3 phases in $\unil$, 3
angles and 3 phases in $\unir$, 3 eigenvalues in $\yuk_D$, and 3
heavy neutrino masses in $M_R$. Once one makes a choice out of the
7 models \textbf{model A} to \textbf{model G} specified in Tables
\ref{table2} and \ref{table3}, one can draw the boundary of
eligible parameter space starting the 6 angles, large or small.
Even though it is supposed that the number of the free parameters
reduces to 12 for 12 equations, considering 6 angles in $|\unir|$
and $|\unil|$ fixed, solving Eq.(\ref{eeseesaw}) yet imbeds a
great deal of ambiguity, since CP phases in MNS matrix and
individual mass eigenvalues in left-hand side can barely be said
as fixed.

There are a few approaches to make improvement in solving the
see-saw mechanism depending on models. First, one can rely on
theories with additional symmetries to reduce the number of
parameters. For example, in case of \textbf{Model G}, a flavor
symmetry like $U(1)$ can be utilized to simplify the mechanism in
Eq.(\ref{eeseesaw}), because the charges of the flavor symmetry
assigned to $\nu_{Ri} \left( \ell_i \right)$ can result in the
common constraint on the side of right (left) mixing in Yukawa
matrix and the structure of right (left) neutrino mass matrix. The
structure of Yukawa matrix in \textbf{Model E} can fit in a theory
of grand-unified-type symmetry between leptons and quarks, if
$\unil_1$ is close to CKM matrix. Among eigenvalues of Yukawa
matrix and 3 heavy masses, there are a few parameters which can be
fixed by hands motivated from a higher-rank symmetry. They can be
$y_3$ of order one and/or $M_3$ of GUT scale. The $y_3 \sim 1$ has
been used in a number of times since Eq.(\ref{ddoffeq2}), as well
as $y_2 \sim y_3$ in particular models.

Another possible approach can be collecting more phenomenological
constraints involved with neutrino Yukawa couplings to diminish
the number of the eligible structure of Yukawa matrix, aiming the
survival of only one natural structure. The classification derived
based on bi-unitary transformations is useful in both approaches.
That is, the models parameterized using bi-unitary transformation
of Yukawa matrix can be easily tested for the compatibility to
additional symmetries as proposed in the previous paragraph. The
bi-unitary parametrization is useful also to examine the
eligibility to additional phenomenology. In describing a process,
the loop contributions of Yukawa couplings consist of a hermitian
operator $\yuk^\dagger \yuk$ or $\yuk \yuk^\dagger$. It is always
possible to rephrase operator in terms of  the minimal number of
parameters, the pair of $(\yuk_D, \unir)$ or the pair of $(\yuk_D,
\unil)$ without any mixing of the two transformations. Thus,
application of an additional constraint will be simply the
re-examination of the 7 models characterized by eigenvalues of
Yukawa matrix, left and right transformations.

\subsection{A model of neutrino masses}

\noindent The description of leptogenesis requires fine
understanding of CP phases in Yukawa matrix as well as CP phases
in light neutrino mass matrix in order to secure a sufficient
amount of CP asymmetry from the decays of heavy Majorana
neutrinos. In lepton sector, however, it is not practical to
regard the CP phases at low-energy as known parameters. Rather, in
analogy of quark sector, the size of a real mixing angle can be
realized as the source of certain amount of CP violation before
the derivation of detailed imaginary parts. Many models proposed
so far took simply order one contribution from CP phase.

As far as the detailed discussion on CP phases is set aside,
\textbf{Models A, D}, and \textbf{E} with a large angle at
$R_{23}$ can provide optimal scenario for the sufficient amount of
CP asymmetry, accordingly lepton asymmetry, Eq.(\ref{aalepto}).
There are a number of cases listed in Table \ref{table1} which can
enhance the CP asymmetry $\epsilon_1$, if $|R_{23}|$ becomes close
to one, so as to reach its maximum value.
\begin{eqnarray}
    \epsilon_1
    \sim 10^{-1} y_3^2 |R_{23}|^2 {M_1 \over M_2}
    \sim 10^{-1} {M_1 \over M_2}.
    \label{eecp}
\end{eqnarray}
It is worthy of attention that large mixing angles in low-energy
originate entirely from the large mixings in heavy neutrinos in
\textbf{Model E}.

The Yukawa matrix in \textbf{Model E} is
\begin{eqnarray}
    |\yuk_N|
    \equiv \left( \begin{array}{ccc}
                1 & \lambda & \lambda \\
                \lambda & 1 & \lambda \\
                \lambda & \lambda & 1 \end{array} \right)
                \left( \begin{array}{ccc}
                    y_1 & 0 & 0 \\
                    0 & 1 & 0 \\
                    0 & 0 & 1 \end{array} \right)
                    \left( \begin{array}{ccc}
                        1 & \rho & \rho \\
                        \rho & 1 & 1 \\
                        \rho & 1 & 1 \end{array} \right),
    \label{eeyukawa}
\end{eqnarray}
where $y_2 \sim y_3 \sim 1,  y_1 \ll y_2$. It is possible that the
\textbf{Model E} reduces  to the model with two-generation of
heavy neutrinos as the large mixing angles become maximal and the
degeneracy in eigenvalues becomes exact. In such limit, the
leading contribution to $\epsilon_1$ does not any longer come from
$R_{23}$. The matrix in Eq.(\ref{eeyukawa}) was introduced in the
basis the mass matrix of righthanded neutrinos is diagonal. Now, a
new basis can be chosen in such a way that Yukawa matrix is
symmetric. Let the Yukawa matrix in \textbf{Model E} in the new
basis be
\begin{eqnarray}
    |\yuk'_N|
    \equiv \left( \begin{array}{ccc}
                1 & \lambda & \lambda \\
                \lambda & 1 & \lambda \\
                \lambda & \lambda & 1 \end{array} \right)
                \left( \begin{array}{ccc}
                    y_1 & 0 & 0 \\
                    0 & 1 & 0 \\
                    0 & 0 & 1 \end{array} \right)
                    \left( \begin{array}{ccc}
                        1 & \rho' & \rho' \\
                        \rho' & 1 & \rho' \\
                        \rho' & \rho' & 1 \end{array} \right).
    \label{eenewbasis}
\end{eqnarray}
Then, the righthanded neutrino mass matrix in Eq.(\ref{eeseesaw})
will have a structure with a large mixing angle in the new basis,
\begin{eqnarray}
   && M'_R \sim \unir' M_R \unir'^T,
    \nonumber \\
   && \unir' \equiv \unir_{new}^* \unir_{old}^T,
    \label{eemnu}
\end{eqnarray}
where $\unir_{old}$ is the right transformation to diagonalize the
Yukawa matrix from the old basis in Eq.(\ref{eeyukawa}), while
$\unir_{new}$ is the one for the diagonalization from the new
basis in Eq.(\ref{eenewbasis}). The \textbf{Model E} can represent
a model in left-right symmetry, in other words, the model shows a
well-balanced correspondence between low energy and high energy
over see-saw mechanism, since the Yukawa matrix is symmetric and
two Majorana neutrino mass matrices are very much of analogy.

The difference of the lepton sector from the quark sector was
recognized from the large mixing angles. In this scenario, the
dissimilarity in the two sectors can be understood as originated
from the existence of massive singlet neutrinos, whereas nothing
like that is in the quark sector. There might have been dynamics
of heavy neutrinos in the early universe to generate the large
mixing angles of which become later the source of the large mixing
angles of light neutrinos. It is shown in Eq.(\ref{eecp}) that the
leptogenesis is optimized. The
Eqs.(\ref{eenewbasis})-(\ref{eemnu}) help the see-saw mechanism
look more meaningful with this model in a theory of left-right
symmetry.

\subsection{Leptogenesis and neutrino large mixing angles}

\noindent Before concluding the section, we give a some thought on
the role of the see-saw mechanism, expressing CP asymmetry in
terms of light neutrino parameters in a model-independent way.
Using solutions to Eq.(\ref{eeseesaw}) for a chosen model, one can
express the CP asymmetry in a proper way to replace the heavy
neutrino masses in terms of low energy neutrino observables and
parameters in Yukawa matrix. The diagonal neutrino mass matrix in
Eq.(\ref{eeseesaw}) can be rephrased as
\begin{eqnarray}
    && m_L = -v_2^2 \mns^T \tilde{\yuk}_N
            M_R^{-1} \mathbb{P}^2 \tilde{\yuk}_N^T \mns,
        \label{eeml}
\end{eqnarray}
where $\tilde{\yuk}_N \equiv \yuk_N \mathbb{P}^\dagger$ is defined
in such a way that phases in $\mathbb{P}$ do not appear explicitly
in $\tilde{\yuk}_N$ as hidden in Eq.(\ref{ccpyyp}). Alternatively,
\bea %
    && M_k \mathbb{P}_k^{-2} = -v_2^2 \sum_{k'} {1 \over m_{k'}}
                       {\left(\mns^T \tilde{\yuk}_N
                       \right)_{k'k}}^2,
    \label{eemr}
\eea %
the imaginary parts of whose inverse are obtained as
\begin{eqnarray}
    && {\sin{2\phi_1} \over M_2} =
    {1 \over v_2^2} \left(\sum_{k} {1 \over m_k}
            {Im \left[\mns^T \tilde{\yuk}_N
             \right]_{k2}}^2 \right)^{-1},\\
    && {\sin{2\phi_2} \over M_3} =
    {1 \over v_2^2} \left(\sum_{k} {1 \over m_k}
            {Im \left[\mns^T \tilde{\yuk}_N
             \right]_{k3}}^2 \right)^{-1},
        \label{eepara}
\eea %
while
\begin{eqnarray}
    M_1 =
    {1 \over v_2^2} \sum_{k} {1 \over m_k}
            {\left[\mns^T \tilde{\yuk}_N
             \right]_{k1}}^2.
    \label{eem1}
\end{eqnarray}
When the value of the rectangular brackets of the above equation
is of order one, the term with $m_1^{-1}$ is guaranteed to be a
significantly leading contribution compared to the other terms
with $m_2^{-1}$ or $m_3^{-1}$ inside the summation, resulting in
the enhancement in the scale of $M_1$. In
Ref.\cite{Berger:2001np}, it was examined that large mixing angles
in MNS matrix can predict the small dilution mass preferred by the
out-of-equilibrium condition in a model. In the right hand side of
Eq.(\ref{ccphase}), $ \sin 2 \phi_1 \left(M_1 / M_2\right)$ and $
\sin 2 \phi_2 \left(M_1 / M_3\right)$ can be re-parameterized only
in terms of Yukawa matrix and light neutrino masses and mixings
using Eqs.(\ref{eepara}) and (\ref{eem1}). The enhancement of
$M_1$ contributed from large mixing angles in Eq.(\ref{eem1})
results in the enhancement in the amount of the CP violation in
Eq.(\ref{ccphase}).

\section{Summary}

\noindent The CP asymmetry and the out-of-equilibrium condition of
lepton flavor violating process are analyzed in terms of
eigenvalues and right mixing angles. In the out-of-equilibrium
condition, there are significant implications. The mixing element
$|\unir_{13}|$ should be small with an upper bound,
Eq.(\ref{ddcondr13}), and two eigenvalues $y_1$ and $y_3$ cannot
be of similar size, Eq.(\ref{ddcondy1}). Depending on which term
in expression of the condition actually leads the condition or
depending on whether eigenvalues are of hierarchy or of similar
size, the magnitude of right transformation is characterized
differently, as shown in Table \ref{table1}. Examining all the
possible models, we found the model independent maximal CP
asymmetry to be $10^{-1} M_1 / M_2$, Eq.(\ref{ccmaxcp}). Besides
thermal out-of-equilibrium condition, we examined also the upper
bound of a LFV decay and mixing angles in neutrino oscillations to
constrain all kinds of parameters, i.e., left and right mixing
elements and eigenvalues. We found that only the seven models can
characterize phenomenologically eligible Yukawa matrices.

The constraint from the LFV did not exclude any choice of
parameter region addition to the category excluded by two other
constraints. In other words, the survival of only seven models is
the outcome of the application of two constraints: the
out-of-equilibrium condition and data of neutrino oscillation. In
Table \ref{table4}, the determination of eligible models from the
application of certain constraints is specified. The bi-unitary
parametrization is useful for model test of the compatibility to
additional symmetries and the eligibility to additional
phenomenology. The \textbf{Model E} is appealing in a sense that
the size of the CP asymmetry is maximized and the number of free
parameters can be reduced based on left-right symmetry.

\begin{acknowledgments}
\noindent The author wishes to express her thanks to Mike Berger
for his advice. She thanks Eung-Jin Chun for the discussion and
suggestion on this work during her stay at KIAS. She thanks
Chuan-Hung Chen since a number of parts were progressed from
discussion with him. She thanks Yasutaka Takanishi for his helpful
comment on LFV.

\end{acknowledgments}

\appendix

\section{Remarks on Tables II and III}

\subsection{Recovery of small angle contribution in transformations}

\noindent In Tables \ref{table2} and \ref{table3}, the elements of
matrices determined from small angles in transformations are not
specified. (They are indicated by a $\ast$.) The \textbf{Models A,
C}, and \textbf{E} include elements, the leading order of which is
first order in $\lambda$ or $\rho$, while other models have their
leading order all zeroth in $\lambda$ and $\rho$. The small values
are not all of the same order of magnitude in a transformation
matrix. However, different small angles do not distinguish because
the comparison of their sizes is not necessary to examine the
eligibility of a model to given constraints in this paper. With
$\rho$'s and $\lambda$'s in $\unir$'s and $\unil$'s, respectively,
the possible symmetric structures of matrix $m_\nu/v_2^2$ of
\textbf{Models A, C,} and \textbf{E} will be as follow;
\begin{widetext} \noindent
\begin{description}
    \item[\textbf{Model A}] \hspace{3pt}
             \bea \left[ \begin{array}{ccc}
              y_1^2 M_1^{-1} & \surd & \surd \\
                \left(y_1^2 M_1^{-1} + y_{23}^2
                M_{23}^{-1}\right)\lambda + y_1 y_{23}M_{123}^{-1}\rho
                & y_{23}^2 M_{23}^{-1} & \surd  \\
                    \left(y_1^2 M_1^{-1} + y_{23}^2
                    M_{23}^{-1}\right)\lambda+ y_1 y_{23}M_{123}^{-1}\rho
                    & y_{23}^2 M_{23}^{-1} & y_{23}^2 M_{23}^{-1}
              \end{array} \right]
              \label{zzmodela} \eea
    \item[\textbf{Model C}] \hspace{3pt}
             \bea \left[ \begin{array}{ccc}
              y_1^2 M_1^{-1} & \surd & \surd \\
              y_i^2 M_i^{-1} \lambda + y_1 \left(y_2 + y_3\right)M_{13}^{-1}\rho &
              y_2^2 M_2^{-1}+y_3^2 M_3^{-1} &
              \surd \\
              y_i^2 M_i^{-1} \lambda + y_1 \left(y_2 + y_3\right)M_{13}^{-1}\rho &
              y_2^2 M_2^{-1}+y_3^2 M_3^{-1} &
              y_2^2 M_2^{-1}+y_3^2 M_3^{-1}
              \end{array} \right] \eea
    \item[\textbf{Model E}]\hspace{3pt}
            \bea \left[ \begin{array}{ccc}
            y_1^2 M_1^{-1} & \surd & \surd \\
            \left(y_1^2 M_1^{-1} + y_2 y_{23} M_{23}^{-1}\right)\lambda
            + y_1 y_2 M_{123}^{-1}\rho &
            y_2^2 M_{23}^{-1} & \surd \\
                \left(y_1^2 M_1^{-1} + y_3 y_{23} M_{23}^{-1}\right)\lambda
                + y_1 y_3 M_{123}^{-1}\rho &
                y_2 y_3 M_{23}^{-1} & y_3^2 M_{23}^{-1}
            \end{array} \right]
            \label{zzmodele}\eea
\end{description} \end{widetext}
where $i=1-3$ and the entry marked by $\surd$ represents the
corresponding symmetric elements.

The 1-2 sector in light neutrino mass needs to be examined to see
whether a model is consistent with for the large mixing angle of
solar neutrino oscillation. In order for a mass matrix to have the
large mixing angle between the first and the second generations,
the 1-2 element of the matrix $m_{\nu12}$ should be not smaller
than the 1-1 element $m_{\nu11}$. If one takes a hierarchical
neutrino mass spectrum $m_2 \sim \sqrt{\triangle m_\odot^2}$ and
$m_3 \sim \sqrt{\triangle m_{atm}^2}$, then the structure of
neutrino mass matrix should require
\begin{eqnarray}
    {m_{\nu12} \over m_{\nu22}}
    \sim \sqrt{\triangle m_\odot^2 \over \triangle
    m_{atm}^2},
\end{eqnarray}
i.e., the ratio between the two elements should be suppressed by
one order of magnitude \cite{Gonzalez-Garcia:2002sm}. The
$m_\nu/v_2^2$ matrices for the three models in
Eqs.(\ref{zzmodela})-(\ref{zzmodele}) show that there can be
parameter region which allows large mixing angle solution for
solar neutrino oscillation as well as the large mixing angles of
atmospheric neutrinos and satisfies the CHOOZ bound of the reactor
neutrinos.

\subsection{Application of constraints}

\begin{table}
\caption{The specification of constraints which in fact rule out
certain types of Yukawa matrices represented by a $\unir$, a
$\unil$, and the comparison of $y_2$ and $y_3$. The C1, C2, C3,
and C4 are the indices of the constraints explained in Section
A.2. For example, Yukawa matrix characterized by $\unir_1,
\unil_1$ and $y_2 \ll y_3$ are forbidden due to the constraint C4.
\vspace{4pt}}
\begin{ruledtabular}
\begin{tabular}{lccccc}
        & & $\unil_1$ \hspace{2pt} & $\unil_2$ \hspace{2pt}
        & $\unil_3$ \hspace{2pt} & $\unil_4$ \\
    \hline
        $\unir_1$ & $y_2 \ll y_3$
        & C4 & C3 & C4 & C3 \\
                  & $y_2 \sim y_3$
                  & \textbf{Model A} \hspace{2pt} & C2 C3 & C2 C3
                  & \textbf{Model E} \\
    \hline
        $\unir_2$ & $y_2 \ll y_3$
        & \textbf{Model B} \hspace{2pt} & C3 & \textbf{Model F} \hspace{2pt} & C3 \\
                  & $y_2 \sim y_3$
                  & C1 & C1 C2 C3 \hspace{2pt} & C1 C2 & C1 C3 \\
    \hline
        $\unir_3$ & $y_2 \ll y_3$ & C4
        & C3 & C4 & C3 \\
                  & $y_2 \sim y_3$ & C1 & C1 C2 C3 \hspace{2pt}
                  & C1 C2  & C1 \\
    \hline\hspace{2pt}
        $\unir_4$ & $y_2 \ll y_3$
        & \textbf{Model C} \hspace{2pt} & C3 & \textbf{Model G} \hspace{2pt} & C3 \\
                   & $y_2 \sim y_3$
                   & \textbf{Model D} \hspace{2pt}
                   & C2 C3 \hspace{2pt} & C2 C3 \hspace{2pt} & C3
    \label{table4}
\end{tabular}
\end{ruledtabular}
\end{table}

\noindent The exclusion of certain types of Yukawa structures is
explained in Table \ref{table4} by specifying the constraint
violated by each particular structure. The constraint C1 is
equivalent to the constraint (i) in listed in Section IV. The
out-of-equilibrium condition of the lepton number violating
process, Eq.(\ref{ddoffeq}), does not allow a structure of Yukawa
matrix with both $y_2 \sim y_3$ and $R_{12} \sim 1$.

The constraint C2 is equivalent to the constraint (ii) in Section
IV. The suppression in $\left( \yuk_N \yuk_N^\dagger \right)_{21}$
for $BR\left( \mu \rightarrow e \gamma \right) < 1.2 \times
10^{-11}$ does not allow the structure of Yukawa matrix with both
$y_2 \sim y_3$ and $L_{12} \sim 1$.

The constraint C3 is equivalent to the constraint (iii) in Section
IV. The phenomenological constraints for light neutrino
experiments require the structure of mass matrix to have $m_{\nu
13} \ll m_{\nu 23} \sim m_{\nu 33}$.

The last constraint C4 is the requirement for Yukawa matrix not to
involve fine tuning. It is not appropriate to assume any possible
fine tuning in Yukawa structure since the comparison of parameters
was done whether it is hierarchical or nearly degenerate, whereas
it is unavoidable to assume  some possible fine tuning in
consideration with light neutrino masses due to the coexistence of
large mixing angles and hierarchical mass spectrum in its
symmetric structure. In the Yukawa matrices, if $y_2 \ll y_3$ when
$L_{23} \sim 1$ and $R_{23} \sim 1$, such a small eigenvalue $y_2$
can be obtained only by fine tuning. For that reason, the cases
with $L_{23} \sim R_{23} \sim 1, y_2 \ll y_3$ are ruled out. In
actual model construction with respect to data, however, the
structures restricted by C4 can be regarded rather positively, if
the fine-tuning is the only way to accommodate data. If one allows
tuning with $L_{23} \sim R_{23} \sim 1, y_2 \ll y_3$, there can be
four more eligible models which are marked by C4 in Table
\ref{table4}.

\end{document}